\documentclass[12pt.a4]{article}

\usepackage{latexsym}
\usepackage{amsmath}
\usepackage{bm}
\usepackage{txfonts}
\title{Covariant Energy Momentum Tensor in General Relativity 
by Generalized Canonical Method}
\author{Masakatsu Kenmoku\thanks{m.kenmoku@cc.nara-wu.ac.jp}\\
Nara Women's University, Nara 630-8506, Japan}
\date{\empty}

\begin{document}
\maketitle
\abstract{
Defining a complete and covariant energy-momentum tensor in general relativity is a longstanding issue of concern. A widely attempted approach is to define it as a pseudo-tensor in an asymptotic Lorentz system.
In this paper, we present a complete covariant definition of the gravitational energy-momentum tensor 
in the generalized canonical form. 

The key is incorporate higher-order differential terms 
of the metric tensor into the gravitational action in order to derive the energy-momentum tensor consistently.
This completely covariant form ensures that the gravitational energy is 
correctly obtained in any coordinate system, 
including polar coordinates and asymptotic Lorentz systems. 
}


\section{Introduction}
\label{intro}

A complete and covariant definition of energy and momentum in general relativity has long been an issue. 
One widely attempted method is to take into account of the quadratic terms of the Christoffel symbol 
in  the gravitational action. 
This approach is to define an asymptotic Lorentzian system as a pseudo-tensor 
as discussed by Einstein, Landau-Lifshitz, M$\phi$ller and others 
\cite{Einstein1916} \cite{Landau1951} \cite{Meller1958} \cite{Misner1973}. 
However, these methods  limit the applicable coordinate systems, 
such as the divergence of vacuum energy in polar coordinates.

We propose a fully covariant energy-momentum tensor for the most fundamental concept  
of gravitational field energy in general relativity to take into account  the surface term consisting of the linear term of the Christoffel symbol  
in addition to the quadratic term of the Christoffel symbol in the gravitational action. 
The surface term contains the second derivative of the metric tensor and the generalized canonical method 
is applied to obtain the fully covariant energy-momentum tensor in general relativity,  
according to the Noether's theorem.

The scalar curvature density, $\sqrt{-g}R$, 
 consists of the square term of the Christoffel symbol $\sqrt{-g}G $ 
and the linear term of the Christoffel symbol $\partial_\mu({\sqrt{-g}W^{\mu}}) $ as a surface term. 
Einstein's equation of motion can be derived from the term $\sqrt{-g}G$  
 and not from the surface term $\partial_\mu({\sqrt{-g}W^{\mu}}) $. 
The energy- momentum tensor, derived from this $ {\sqrt{-g}G}$ term, is no longer covariant under general coordinate transformations 
as a pseudo-tensor, and its applicability is limited.

We take into account for the surface term $ \partial_\mu({\sqrt{-g} W^{\mu}})$, 
which includes the second derivative of the metric, 
in addition to the ${\sqrt{-g}G}$ term.
This term $ \partial_\mu({\sqrt{-g} W^{\mu}})$ does not contribute to the equation of motion, due to the surface term.  
However, it contributes to the gravitational energy-momentum tensor,  
making it fully covariant under general coordinate transformations. 

Note that,
1) the second derivative of the metric
$ {g_{\mu\nu}}$ 
is included in a surface term, and therefore does not contribute to the equation of motion. 
Therefore, no ghosts occur either.
2) In addition to 
${\delta{g}_{\mu\nu}=0}$ on ${\partial{\Sigma}}$,
we impose 
${\delta{g}_{\mu\nu}=0}$ on ${\partial^{2}{\Sigma}}$ 
as boundary conditions in the variation principle for the entire spacetime
$ {\Sigma}$.
The conservation of the resulting covariant energy-momentum tensor can be shown by direct calculation.
The total energy-momentum tensor obtained in this way is invariant under general coordinate transformations, and it gives the same value not only in asymptotic Lorentzian systems but also in polar coordinates and any other system.

This paper is organized as follows: 
Section 2 derives the energy-momentum tensor in generalized canonical form in general relativity. 
Section 3 provides examples of applications of gravitational field energy to vacuum spacetime in 3.1 and Schwarzschild black hole 
in 3.2. Section 4 provides conclusions and a discussion. 
The appendix presents a simple scalar field model to help to understand higher-order derivative theories.


\section{Generalized Canonical Method for the Energy-Momentum Tensor in General Relativity}
\renewcommand{\theequation}{\thesection,\arabic{equation}}
\setcounter{equation}{0}

This chapter derives the generalized canonical form for the energy-momentum tensor in general relativity. The canonical method is generalized to include second derivatives.

The total action consists of the gravitational Einstein-Hilbert action and the matter action, 
\begin{equation}
I = I_{g}+I_{m} =\int _{\Omega}L_{g}\sqrt{-g}d^4 x + \int _{\Omega}L_{m}\sqrt{-g}d^4 x ,
\end{equation}
where $\Omega$ denotes the integration spacetime region.

The gravitational Lagrangian can be divided into two terms: one consisting of the square of the first derivative 
of the metric tensor $\textbf{G}$ and the other consisting of the second derivative of the metric tensor $\textbf{W}$.
\begin{equation}
L_g = \frac{1}{2\kappa} R = L_{G} + L_{W}, 
\end{equation}
where $R$ denotes the scalar curvature and the Einstein constant of gravitation  
$\kappa=8\pi G/c^4=8\pi$  is used with the convention $c=G=1$. 
The first part of the Lagrangian  $L_{G}$ is expressed as the product of two Christoffel symbols 
$\Gamma^{\lambda}_{\ \mu\nu}=\frac{1}{2}g^{\lambda\sigma}(g_{\nu\sigma,\mu}+g_{\mu\sigma, \nu}-g_{\mu\nu, \sigma})$ 
as follows: 
\begin{eqnarray}
\textbf{L}_{G} :=\sqrt{-g}L_{G}
= \frac{1}{2\kappa}\textbf{G},  \, \, 
\textbf{G}:=\sqrt{-g}G
 = \sqrt{-g} g^{\mu\nu}
(\Gamma^{\lambda}_{\mu\rho}\Gamma^{\rho}_{\nu\lambda}
-\Gamma^{\lambda}_{\mu\nu}\Gamma^{\rho}_{\lambda\rho}) ,
\end{eqnarray}
where the boldface notation for the density $\textbf{O}=\sqrt{-g}O$ is used. 
The second part of the Lagrangian $L_{W}$ is expressed as the first-order divergent term of the Christoffel symbol 
as in the surface term:
\begin{eqnarray}
\textbf{L}_{W}:=\sqrt{-g}L_{W} = \frac{1}{2\kappa} \textbf{W} ,
\,\, \textbf{W}:= \partial_{\mu}\textbf{W}^{\mu}:=\partial_{\mu} \left\{\sqrt{-g} (g^{\alpha\beta} 
\Gamma^{\mu}_{\alpha\beta} -
 g^{\mu\nu} \Gamma^{\lambda}_{\nu\lambda}) \right\},
\end{eqnarray}

In this section, we will derive the equations of motion and conserved quantities for each of $ L_{G}$ and $L_{W}$. 
According to the Noether's theorem \cite{Noether1918},  
we will obtain the total energy-momentum tensor of the gravitational field 
in covariant form.


\subsection{The Gravitational Energy-Momentum Tensor for the $\textbf{G}$ Term} 

We  consider the variation of the $\textbf{G}$ term contribution in the action in eq.(2.3) as 
\begin{align}
	\delta{I_{G}}
&=\delta \left\{ \frac{1}{2\kappa} \int \textbf{G}d^4 x \right\} 
=\frac{1}{2\kappa}\int \left\{ \bar{\delta}(\textbf{G}) + \partial_\lambda (\textbf{G} \delta x^{\lambda}) \right\} d^4 x 
\nonumber\\
&=\frac{1}{2\kappa}\int \left\{ -(R^{\mu\nu}-\frac{1}{2}g^{\mu\nu}R) \sqrt{-g}\bar{ \delta} g_{\mu\nu}
+ \partial_{\lambda} (\frac{\partial \textbf{G}}{\partial g_{\mu\nu, \lambda}}
\bar{\delta} g_{\mu\nu} 
+\textbf{G}\delta x^{\lambda}) 
\right\} d^4x ,
\end{align}
with the derivative  notation $O_{\lambda}=\partial_{\lambda}O=\partial O/\partial x^{\lambda} $. 
The variations $\delta$ and $\bar{\delta} $ for a field $\phi(x)$ are defined as 
\begin{equation}
\delta \phi (x) :=\phi' (x')-\phi(x) \, \, \, \text{and} \, \, \,  
\bar{\delta} \phi (x) :=\phi' (x)-\phi(x) \backsimeq \delta{\phi}(x) -\delta x^{\mu} \partial_{\mu}\phi(x) , 
\end{equation}
with the property $\bar{\delta}(\partial_{\mu} \phi)=\partial_{\mu}(\bar{\delta} \phi)$. 

The variation of the action leads to the Einstein's equation under the condition 
on the boundary surface  $\bar{\delta}g_{\mu\nu}=0$ and  $\delta x^{\mu }=\text{constant}$ on  $\partial \Omega$: 
\begin{equation} 
G^{\mu\nu}=R^{\mu\nu}-\frac{1}{2}g^{\mu\nu}R=\kappa T_{sym}^{\ \mu\nu} , 
\end{equation}
with the matter symmetric energy-momentum tensor
\begin{equation}
\textbf{T}_{sym}^{\mu\nu}=\sqrt{-g}T_{sym}^{\mu\nu}=2\frac{\partial {\textbf{L}}_{m}} {\partial g_{\mu\nu}} .
\end{equation}

The canonical gravitational energy-momentum tensor 
${\textbf{t}}_{G \ \nu}^{\ \mu}$ 
for the $\textbf{G}$-term is defined as follows: 
\begin{equation}
\textbf{t}_{G\, \,  \nu }^{\ \mu}
=\frac{1}{2\kappa} (\frac{\partial \textbf{G}}{\partial g_{\alpha\beta, \mu}} g_{\alpha\beta , \nu}
-\delta^{\mu}_{\nu}\textbf{G})., 
\end{equation}
and the conservation law is obtained using the Einstein's equation and the matter field equation, 
\begin{equation} 
\partial_{\mu}  ( \textbf{t}_{G \, \nu}^{\ \mu} +\textbf{T}_{can\, \nu}^{\ \mu} )  =0., 
\end{equation}
where the matter canonical energy-momentum tensor is defined: 
\begin{equation}
\textbf{T}_{can\, \nu}^{\ \mu}:=\frac{\partial \textbf{L}_m}{\partial \phi_{, \mu}}\phi_{ , \nu} 
-\delta^{\mu}_{\nu} \textbf{L}_m , 
\end{equation}

 Note that this energy-momentum tensor satisfies the conservation law  in eq.(2.10)  
but it is not generally covariant as the pseudo energy-momentum tensor.   


\subsection{The Gravitational Energy-Momentum Tensor for the $\textbf{W}$ Term }

In this subsection, we consider the $\textbf{W}$ term defined in eq.(2.4). This term does not contribute to the 
equation of motion but to the energy-momentum tensor.  
 
The variation the corresponding action is evaluated in detail as follows,
\begin{align}
\bar{\delta} I_W  
&= \bar{\delta}  \left\{ \int_{\Omega}\frac{1}{2\kappa} \textbf{W}d^4x \right\} \nonumber \\
&=\frac{1}{2\kappa} \int_{\Omega} \left\{  \bar{\delta} ( \textbf{W}) 
+\partial_{\lambda}(\textbf{W} \delta x^{\lambda} ) \right\} d^4 x \nonumber \\
&=\frac{1}{2\kappa} \int_{\Omega} \left\{  
\frac{\partial  \textbf{W}} {\partial g_{\alpha\beta} }\bar{\delta}g_{\mu\nu}
+\frac{\partial  \textbf{W}} {\partial g_{\alpha\beta, \gamma} } \bar{\delta} g_{\alpha\beta, \gamma}
+\frac{\partial  \textbf{W}^{\mu}} {\partial g_{\alpha\beta, \gamma\lambda} }
\bar{\delta} g_{\alpha\beta,\gamma\lambda}
+\partial_{\lambda}(\partial_{\mu}\textbf{W}^{\mu} \delta x^{\lambda} ) \right\} d^4 \nonumber \\
&=\frac{1}{2\kappa} \int_{\Omega} \left\{  
(\frac{\partial  \textbf{W}} {\partial g_{\alpha\beta} }
-\partial_{\gamma}\frac{\partial  \textbf{W}} {\partial g_{\alpha\beta, \gamma} } 
+\partial_{\gamma}\partial_{\lambda}\frac{\partial  \textbf{W}} {\partial g_{\alpha\beta, \gamma\lambda} })
\bar{\delta} g_{\alpha\beta} +
\partial_{\gamma}(\frac{\partial  \textbf{W}} {\partial g_{\alpha\beta, \gamma} } \bar{\delta}g_{\alpha\beta}) 
\right. \nonumber \\ 
&\ \ \ \ \ \ \ \ \ \ \ \ \ \ \ \ \left. 
-2
\partial_{\lambda}(\partial_{\gamma}(\frac{\partial\textbf{W}}{\partial g_{\alpha\beta, \gamma\lambda}})
\bar{\delta}g_{\alpha\beta})
+\partial_{\gamma}\partial_{\lambda}(\frac{\partial  \textbf{W}^{\mu}} {\partial g_{\alpha\beta, \gamma\lambda} }
\bar{\delta} g_{\alpha\beta}) 
+\partial_{\lambda}(\textbf{W} \delta x^{\lambda} ) \right\} d^4 x  .
\end{align}
The equation of motion is obtained under the condition of 
$\partial g_{\mu\nu}=0$ with  $\delta x^{\mu }=\text{constant}$ 
on $\partial \Omega$ and also $\partial \partial \Omega$, and it is shown to be identically zero: 
\begin{align}
\frac{\partial  \textbf{W}} {\partial g_{\alpha\beta} }
-\partial_{\gamma}\frac{\partial  \textbf{W}} {\partial g_{\alpha\beta, \gamma} } 
+\partial_{\gamma}\partial_{\lambda}\frac{\partial  \textbf{W}} {\partial g_{\alpha\beta, \gamma\lambda} }=0 .
\end{align}
The result is understandable because of its surface-boundary term.  
The gravitational energy-momentum tensor is obtained from eqs. (2.12) and (2.13) as
\begin{align}
\textbf{t}_{W\, \, \nu}^{\mu}
=\frac{1}{2\kappa}\left\{
(\frac{\partial  \textbf{W}} {\partial g_{\alpha\beta, \mu} }  
-2\partial_{\gamma} 
\frac{\partial\textbf{W}}{\partial g_{\alpha\beta\gamma\mu}})
g_{\alpha\beta \, , \nu}
+\partial_{\gamma}(\frac{\partial  \textbf{W}} {\partial g_{\alpha\beta, \gamma\mu} }g_{\alpha\beta \, , \nu})
- \delta^{\mu}_{\nu}\textbf{W} 
\right\}
\end{align}
which satisfies the conservation law: 
\begin{align}
\partial_{\mu} \textbf{t}_{W\, \nu}^{\ \mu}=0 ,
\end{align}
as expected by the Noether's theorem and is also confirmed by the direct calculation. 
Note that 
the generalized canonical method is applied to treat the second derivative of the metric tensor $g_{\mu\nu , \alpha\beta}$
in deriving the gravitational energy-momentum tensor
$ \textbf{t}_{W \ \nu}^{\ \mu}$. 

The contribution of the $\textbf{W}$ is very important for making the energy-momentum 
to be fully covariant under the general coordinate transformations. 
The appendix presents a simple scalar model to help understand the higher-order derivative theory. 

\subsection{Total Covariant Energy-MomentumTensor}

The total gravitational energy-momentum tensor is obtained by adding the $\textbf{t}_{G}$ in eq.(2.9) 
and $\textbf{t}_{W}$ in eq.(2.14) , 
and then expressing in terms of the Ricci scalar curvature density $\textbf{R}=\sqrt{-g}R$ as follows: 
\begin{align}
\textbf{t}^{\mu}_{\, \, \nu}
&=\textbf{t}_{G\, \nu}^{\ \mu}+\textbf{t}_{W\, \nu}^{\ \mu} \nonumber \\
&=\frac{1}{2\kappa} \left\{(\frac{\partial \textbf{G}}{\partial g_{\alpha\beta, \mu}} g_{\alpha\beta ,\nu}
-{\delta}^{\mu}_{\nu}\textbf{G})
+
(\frac{\partial  \textbf{W}} {\partial g_{\alpha\beta, \mu} }  
-2
\partial_{\gamma} \frac{\partial\textbf{W}}{\partial g_{\alpha\beta, \gamma\mu}})g_{\alpha\beta , \nu}
+\partial_{\gamma}(\frac{\partial  \textbf{W}} {\partial g_{\alpha\beta, \gamma\mu} }g_{\alpha\beta , \nu})
 - \delta^{\mu}_{\nu}\textbf{W} \right\}
\nonumber \\
&=\frac{1}{2\kappa} \left\{
(\frac{\partial  \textbf{R}} {\partial g_{\alpha\beta, \mu} }  
-2
\partial_{\gamma} \frac{\partial\textbf{R}}{\partial g_{\alpha\beta, \gamma\mu}})g_{\alpha\beta , \nu}
+\partial_{\gamma}(\frac{\partial  \textbf{R}} {\partial g_{\alpha\beta, \gamma\mu} }g_{\alpha\beta , \nu})
 - \delta^{\mu}_{\nu}\textbf{R} \right\}. 
\end{align}
We used the fact that the second derivatives of metric tensor $g_{\alpha\beta,\gamma\mu}$
are included only in the $\textbf{W}$ and the $\textbf{R}$ terms. 

As a result, the total gravitational energy-momentum tensor plus the matter canonical energy-momentum tensor 
is fully covariant and conserves by the logical consideration and by the direct calculation 
as in eqs.(2.10) and (2.15): 
\begin{align}
\partial_{\mu}(\textbf{t}^{\mu}_{\ \nu}+\textbf{T}_{can \ \nu}^{\ \mu})=0 . 
\end{align}

In the next section, we will consider the applications 
to understand the full covariance under the general coordinate transformations. 

\section{Application}

We have derived the gravitational energy-momentum tensor density in its fully covariant form.  
Gravitational energy is contributed from the $\textbf{G}$ and $\textbf{W}$ terms 
in eqs.(2.3) and (2.4) as 
$
\textbf{L}_G=\frac{1}{2\kappa}\textbf{R}=\frac{1}{2\kappa}(\textbf{G}+\textbf{W}) 
$ , 
where $\textbf{R}$ is the Ricci scalar, and is expressed as 
\begin{align}
\textbf{t}^{\ 0}_{\ \ 0}=\textbf{t}^{\ 0}_{\ G\ 0}+\textbf{t}^{\ 0}_{W\ 0} ,
\end{align}
where in the generalized canonical form 
\begin{align}
\textbf{t}^{\ 0}_{\ G\ 0}&=\frac{1}{2\kappa}(\frac{\partial \textbf{G}}{\partial g_{\alpha\beta, 0}}g_{\alpha\beta , 0} 
-\delta^{0}_{0}\textbf{G}) \ \ \  \text{and} \nonumber \\
\textbf{t}^{\ 0}_{\ W\ 0}&=\frac{1}{2\kappa}\left\{(\frac{\partial \textbf{W}}{\partial g_{\alpha\beta, 0}}
-2\partial_{\gamma} \frac{\partial\textbf{W}}{\partial g_{\alpha\beta , \gamma 0}})g_{\alpha\beta , 0} 
-\partial_{\gamma} (\frac{\partial\textbf{W}}{\partial g_{\alpha\beta, \gamma 0}} g_{\alpha\beta , 0} )
-\delta^{0}_{0}\textbf{W} \right\} .
\end{align}
The gravitational energy is given by the volume integral of the energy density: 
\begin{align}
E_g= -\int \textbf{t}^{0}_{\ 0} d^3x=-\int (\textbf{t}_{G\ 0}^{0} +\textbf{t}_{W\ 0}^{0})\ d^3x , 
\end{align}
where we take into account the positivity of the energy 
as $-\textbf{t}^{0}_{\ 0}{\geq 0}$.  

Simple and important applications are presented to the vacuum spacetime in subsection 3.1 and 
the Schwarzschild black hole in subsection 3.2.  

\subsection{Application to Vacuum Spacetime} 

Simple but important application to the vacuum spacetime is evaluated in the polar and 
the Cartesian coordinate systems. 

\subsubsection*{A)  In  the  polar coordinate system: } 
The invariant distance is
\begin{align}
ds^2=-dt^2+dr^2+r^2(d\theta^2+\sin \theta^2d\phi^2). 
\end{align}
The contribution of the gravitational energy to the vacuum spacetime is evaluated from 
$\textbf{G}$ and $\textbf{W}$ terms respectively,  
\begin{align}
\textbf{G}=2\sin \theta \ \ \ \text{and} \ \ \ \textbf{W}=-2\sin{\theta} . 
\end{align}
The volume integral for each term gives an infinite value as
\begin{align}
-\int_{\partial \Omega} \textbf{t}_{G\ 0}^{0} d^3 x=\frac{1}{2\kappa}\int_{\partial{\Omega}} \textbf{G} d^3 x
=\frac{1}{\kappa}\int_{\partial \Omega} r^2 \sin^2 \theta dr d\theta d\phi 
=\frac{\pi^2r^3}{3\kappa} \big{|}^\infty \rightarrow \infty.,
\end{align}
and similarly
 \begin{align}
-\int_{\partial \Omega} \textbf{t}_{W\ 0}^{0}d^3 x=\frac{1}{2\kappa}\int_{\partial{\Omega}} \textbf{W} d^3 x
=\frac{-1}{\kappa}\int_{\partial \Omega} r^2 \sin^2 \theta dr d\theta d\phi 
=\frac{-\pi^2r^3}{3\kappa} \big{|}^\infty \rightarrow -\infty.
\end{align}
The sum of these values is zero. 
This fact shows the importance of the $\textbf{W}$ term. 

\subsubsection*{ B)  In the flat orthogonal Cartesian coordinate system:}
The invariant distance is given by, 
\begin{align} ds^2=-dt^2+ dx^2+dy^2+dz^2 .
\end{align}
Each contribution from the $\textbf{G}$ and $\textbf{W}$ terms is zero, so the total gravitational energy is of course zero. 
\subsection{Application to the Schwarzschild Black Hole} 

The next subsection presents very important and interesting application to the Schwarzschild black hole. 

\subsubsection*{A) In the polar coordinate system:}
The standard expression of the Schwarzschild black hole is 
\begin{align}
ds^2=-(1-\frac{2m}{r}) dt^2+(1-\frac{2m}{r})^{-1} dr^2+r^2(d\theta^2+\sin \theta^2d\phi^2), 
\end{align}
where $m$ is the mass of the matte with the convention $c=G=1$.  

The gravitational energy comes from eqs. (2.16) and (2.20) as
\begin{align}
E_g=-\int \textbf{t}^{0}_{\ 0} d^3x=\frac{1}{2\kappa}\int \textbf{R} d^3 x. 
\end{align} 

To estimate the value of $\textbf{R}$, we consider the Einstein's tensor 
$G^{\mu}_{\ \nu}=R^{\mu}_{\ \nu}-\frac{1}{2}\delta^{\mu}_{\ \nu}R$
and take the trace to get the relation $R=-G^{\mu}_{\ \mu}$. 

To avoid the ambiguity of the singular point in the Schwarzschild solution, 
we apply the regularization method to the factor $X:=1-2m/r$ as 
\begin{align}
	X:_{\epsilon}=1-\frac{2m}{r_{\epsilon}} \ \ , \ \ r_{\epsilon}:=\sqrt{r^2+\epsilon^2} , 
\end{align}
where a small regularization parameter $\epsilon$ is used. 
Then we estimate each term of the Einstein tensor:  
\begin{align}
G^0_{\ 0}=G^1_{\ 1}=-(\frac{1}{r^2}-\frac{X_{\epsilon}}{r^2}-\frac{{X^,_{\epsilon}}}{r})=-\frac{2m \epsilon^2}{r^2r_{\epsilon}^3} , 
\end{align}
and 
\begin{align}
G^2_{\ 2}=G^3_{\ 3}=\frac{1}{2}(X''_{\epsilon} + \frac{2}{r}X'_{\epsilon}) =\frac{1}{2}\Delta X_{\epsilon} = - m \Delta \frac{1}{r_{\epsilon}} . 
\end{align}
The gravitational energy is obtained by the volume integration of each term of the Einstein tensor. 
For $G^0_{\ 0}+G^1_{\ 1}$ in eq.(3.29) , we have
\begin{align}
  \int (G^0_{\ 0}+G^1_{\ 1})\sqrt{-g} d^3 x=-16\pi m . 
\end{align} 
For $G^2_{\ 2}+G^3_{\ 3}$ , the naive calculation tends to zero as $\epsilon \rightarrow 0$ but the integration gives a finite contribution 
\begin{align}
\int (G^2_{\ 2}+G^3_{\ 3})\sqrt{-g}d^3 x=8\pi m. 
\end{align}
Note that the singular behavior, i.e., zero or $8\pi m$,  of the Laplacian operator in the limit $\epsilon \rightarrow 0$
and the delta function is the source of the Laplacian operator: 
\begin{align}
\Delta \frac{1}{r}=-4\pi \delta^{(3)}(\bf{r}). 
\end{align}
The origin of this discrepancy is the contribution from the matter source located at the origin, 
which is the outside of the gravitational field,.

Therefore, the gravitational energy for the Schwarzschild black hole is given by 
\begin{align}
E_g&=-\int \textbf {t}^{0}_{\ 0}d^3 x = \frac{1}{2\kappa}\int \textbf{R}d^3 x \nonumber \\
&=-\frac{1}{2\kappa} \int(G^0_{\ 0}+G^1_{\ 1}+G^2_{\ 2}+G^3_{\ 3})\sqrt{-g} d^3 x 
= \frac{1}{2\kappa}(16\pi m - 8\pi m) =\frac{m}{2} , 
\end{align}
and the canonical matter energy
\begin{align}
E_m=\frac{1}{2\kappa}\int 8\pi m\delta^{(3)}({\bf{r}}) d^3 x=\frac{m}{2} .
\end{align}
The sum of these terms gives the total Schwarzschild black hole energy:
\begin{align}
E=E_g +E_m = \frac{m}{2}+\frac{m}{2}=m. 
\end{align} 

\subsubsection*{ B) In the isolated coordinate system as asymptotic Cartesian coordinate:}

The standard Schwarzschild black hole solution in eq.(2.26) is rewritten using the rectangular variables: 
\begin{align}
ds^2=-(1-\frac{2m}{r})dt^2 + \sum_{k=1}^3 (dx^k)^2 +\frac{2m}{r^2(r-2m)}(\sum_{k=1}^3 x^kdx^k) .
\end{align}
The asymptotic behavior of this metric is 
\begin{align}
ds^2 \rightarrow  -(1-\frac{2m}{r}) dt^2+ (1+\frac{2m}{r})(dx^2+dy^2+dz^2) \ \ \text{as} \ \  r\rightarrow \infty
\end{align}
and approaches to the Minkowskian spacetime, which is the isolated coordinate system.  

The gravitational energy is estimated for the isolated frame 
for the $\textbf{G}$ and $\textbf{W}$ terms respectively.  
Using the approximate values of the Christoffel symbol for $r\rightarrow \infty$, we have
\begin{align}
\Gamma_{mn}^{ k} &\approx \frac{m}{r^3}(\delta_{mn} x^k-\delta_{km}x^n-\delta_{kn}x^m) ,  \nonumber \\ 
\Gamma_{00}^{ k} &\approx \frac{m}{r^3} x^k , \ \ \ 
\Gamma_{0n}^{ 0} \approx \frac{m}{r^3} x^n , \ \ \ \text{others} \approx O\ ( 1/r^3 ) , 
\end{align}
the \textbf{G} term contribution in eq.(2.3) is estimated as 
\begin{align}
\textbf{G}=\sqrt{-g}g^{\mu\nu}(\Gamma_{\rho\tau}^{ \lambda}\Gamma_{\nu\lambda}^{ \tau}
-\Gamma_{\mu\nu}^{ \lambda}\Gamma_{\lambda\tau}^{ \tau}) \approx O\ (2m/r^2)^2 , 
\end{align}
and the contribution is negligible in this frame. 

The $\textbf{W}^{\mu}$ term in eq.(2.4) is estimated using the approximate expression of the Christoffel symbol in eq.(3.39)  
\begin{align}
\textbf{W}^{\mu}=\sqrt{-g}(g^{\alpha\beta}\Gamma_{\alpha\beta}^{ \mu} -g^{\mu\nu} \Gamma_{\nu\lambda}^{\lambda}) 
\simeq \sqrt{-g} g^{\mu\nu} g^{\alpha\beta}(\partial_{\alpha}g_{\nu\beta}-\partial_{\nu}g_{\alpha\beta}) , 
\end{align}
and the \textbf{W} contribution is obtained 
\begin{align}
\textbf{W}
&=\partial_{\mu}\textbf{W}^{\mu}= \partial_k \textbf{W}^k 
\approx  \partial_k \left\{  \sqrt{-g}g^{kl}g^{\alpha\beta}(\partial_{\alpha} g_{l\beta}-\partial_l g_{\alpha\beta}) \right\}
\nonumber\\
&= \partial_k \left\{  \sqrt{-g}g^{kl}g^{00}(\partial_{\alpha} g_{l\beta}-\partial_l g_{\alpha\beta}) 
+\sqrt{-g}g^{kl}g^{\alpha\beta}(\partial_{\alpha} g_{l\beta}-\partial_l g_{\alpha\beta})\right\} \nonumber \\
&\simeq\partial_k \left\{  (\partial_k g_{00} -\partial_0g_{k0})+ (\partial_l g_{kl} -\partial_kg_{ll}) \right\} , 
\end{align}
here the Greek indices $\mu, \nu, , , ,=0,1,2,3$ and the Latin indices $k, l, ... =1,2,3 $ are used. 
In the isolated system in eq.(3.38), the metrics are expressed approximately  
\begin{align}
&g_{\mu\nu}(x) \simeq \eta_{\mu\nu}+h_{\mu\nu}(x) , \nonumber \\
& \eta_{\mu\nu}=\text{diag}(-1,1,1,1) 
\ \ \ \text{and}\ \ \ h_{kl}(x)\simeq \frac{2m}{r^3}x^kx^l
\end{align}
and their derivatives are 
\begin{align}
h_{kl,l}\simeq \partial_k (\frac{2m}{r^3}x^kx^l)=\frac{2m}{r^3}x ^k \ \ \ \text{and} \ \ \ 
h_{ll,k} \simeq \partial_k (\frac{2m}{r^3}(x^l)^2)=\partial_k\frac{2m}{r} . 
\end{align}
The approximate value of $\textbf{W}$ term is calculated summarizing them as
\begin{align}
\textbf{W}\simeq\partial_k(\partial_k g_{00}-\partial_0 g_{k0}) +\partial_k(\partial_lg_{kl} -\partial_kg_{ll})
=\partial_k^{\ 2}(\frac{2m}{r})+\partial_k(\frac{2m}{r^3}x^k-\partial_k \frac{2m}{r})
\end{align}
and the gravitational energy is obtained 
\begin{align}
E_g=\frac{1}{2\kappa} \int \textbf{R} d^3 x \simeq \frac{1}{2\kappa} \int \textbf{W} d^3 x 
=-\frac{m}{2} +m =\frac{m}{2} . 
\end{align}
On the other hand, the canonical matter energy has the same value in the polar coordinate system 
because of the gauge invariance $E_m=m/2$ in eq.(3.36). Then the total energy is
\begin{align}
E=E_g +E_m=\frac{m}{2}+\frac{m}{2}= m, . 
\end{align} 
as expected.

We note that the relation between our derivation of the gravitational energy and the ADM energy \cite{ADM1959}, 
which appears in the special part of the metrics in eq.(3.42) as 
\begin{align}
\frac{1}{2\kappa}\int \textbf{W} dx^3 \big|{\text{special part}}
= \frac{1}{2\kappa} \int \partial_k  (\partial_l g_{kl} -\partial_kg_{ll}) dx^3
=\frac{1}{2\kappa}\int (\partial_l g_{kl} -\partial_kg_{ll})d\sigma_k =m . 
\end{align}  

We showed that the Schwarzschild black hole energy, which is the sum of the gravitational energy and the canonical 
matter energy, has the same value in the standard polar coordinate within an isolated system. In this system, the 
main  contribution comes from the $\textbf{W}$ term, not from the $\textbf{G}$ term, 
which is usually considered mainly.  
The ADM energy also included in the \textbf{W} term in our generalized canonical method. 

\section{Conclusions and Discussion}

We derived the gravitational energy-momentum tensor by the generalized canonical method 
in fully covariant form under the general coordinate transformation using the generalized canonical method.  
In this method, we take into account of 
the second-order derivative terms of the metric tensor as in the \textbf{W} term in addition to the product 
of the first-order derivatives of the metric in the $\textbf{G}$ term, which is considered by other research works 
up to this point.  
In usual variation method, the second-order derivative terms of the metric tensor is not under consideration.  
We extended the boundary condition to include as
\begin{align}
\delta g_{\mu\nu}(x)=0  \ \ \text{on} \ \ \partial \Omega \ \ \text{and} \ \ \partial \partial \Omega ,
\end{align}
and the energy contribution from the $\textbf{W}$ term is obtained by using of the extended application of Noether's theorem. 
 
The gravitational energy obtained from the $\textbf{W}$ term is confirmed to be conserved through the logical consideration 
as well as the direct calculation.  
Consequently, the total gravitational energy together, including both the $\textbf{G}$  
and the $\textbf{W}$ terms, is invariant under the general coordinate transformation. 

As an application, we consider the vacuum spacetime and the Schwarzschild black hole  
in the polar coordinate  and the asymptotic Cartesian coordinates, 
and obtain the same total energy values. 
In case of vacuum spacetime, both of the $\textbf{G}$ and $\textbf{W}$ terms are infinity contribution 
with the opposite sign and the sum of them is zero  in the subsection 3.1. 
In the case of the Schwarzschild black hole, the $\textbf{G}$ term is considered negligible, 
and the $\textbf{W}$ term is shown to be main contribution to the energy, 
 in the asymptotic Cartesian coordinate in the subsection 3.2,
\\

 Some discussion is taken into account. \\

1) The ADM energy is expressed by adding two terms in the asymptotic Cartesian coordinate, which is  
expressed by using the Gauss's law \cite{ADM1959},  
\begin{align} 
E_{ADM} (1st) &\simeq \frac{1}{16\pi } \oint_{S^2_{\infty}} d\Omega^2_2 r^2 
\left\{ (\partial_r g_{00}-\partial_0g_{r0})  + r^k( \partial_i g_{ki} -\partial_k g_{ii})\right\} , \nonumber \\
E_{ADM} (2nd) &\simeq - \frac{1}{16\pi } \oint_{S^2_{\infty}} d\Omega^2_2 r^2 
(\partial_r g_{00}-\partial_0g_{r0})  \nonumber \\
E_{ADM}&=E_{ADM}(1st)+E_{ADM}(2nd) 
\end{align}
  
The ADM energy  $E_{ADM}(1st)$, which is known as the Komar integral \cite{Komar1959},  
corresponds to and coincides with the $\textbf{W}$ contribution in eq.(3.42) 
to give the contribution $m/2$. 
The ADM energy $E_{ADM}(2nd)$ corresponds to the matter canonical energy to give $m/2$. 
The total ADM energy is equal to our covariant gravitational energy plus the matter canonical energy as $m$
corresponding to eq.(3.47). 

It is worthwhile noting that the ADM energy method can be applied to the asymptotic Minkowskian spacetime 
while our energy is in a fully covariant form and can be applied to any space-time. \\

2) There are some trials for the derivation on the conservation of the gravitational energy by the Noeter's second theorem 
\cite{Noether1918}. 
This method derives the quasi-local energy, which is not the tensor 
but the pseudo-tensor under the general coordinate transformation \cite{Utiyama1984}.  
On the other hand, our generalized canonical method derives fully covariant energy-momentum tensor.  \\

3) Recently, 
Aoki et.al. published a series of papers on the conserved energy and entropy in general rerativity \cite{Aoki2021} 
\cite{AokiOnogi2022} \cite{AokoOnogiYamaoka2024}. 
The review article by Aoki appeared as the 40th anniversary special issues of 
International Journal of Modern Physics and Modern Physics Letters \cite{Aoki2025}.  
They introduced the Killing vector $\xi^{\mu}$ to the Einstein's equation of the gravitational field to evaluate the 
symmetric energy of matter for the Schwarzschild black hole and others.  
They estimate the symmetric energy of matter by the Einstein tensor using the distribution technique. 

On the other hand, we calculate the gravitational energy using the general canonical method, which is 
fully covariant under coordinate transformation and not restrict to using the Killing vector.  
\\

Note added: This research work was presented at the XVI International 
Conference on Gravitation, Astrophysics and Cosmology (ICGAC16) 
held at SYSU Academic Vila, Shenzhen, China, Aug 10-16 2026.

\section*{Acknowledgements} 

I would like to thank to Professor Kazuyasu Shigemoto for his insightful discussions and valuable comments, 
especially regarding to the vacuum calculation in the polar coordinate system.  
The author would also like to thank to Professor Shinya Aoki for his excellent research works 
on the gravitational energy related to our research theme. 

\section*{Appendix: The Higher Order Derivative Scalar Field Model} 

To help understand the theory including the second order derivative field 
as the $\textbf{W}$ term in subsection 2.2, 
the scalar field model is considered. 
The scalar field Lagrangian $\Delta L_{\phi}$ is assumed to be in the following form: 
\begin{align*}
\Delta L_{\phi}=\partial_{\alpha}(\eta^{\alpha\beta}\phi \phi_{, \ \beta})
=-\partial_t (\phi \dot{\phi})+ \partial_x (\phi \phi')
\end{align*}
where we have used the notation $\eta^{\alpha\beta}=(-1,1)$, $\dot{\phi}:=\partial_t \phi$ and $\phi':=\partial_x\phi$. 
The variation for the action of $\Delta L_{\phi}$  corresponding to the calculation in eq.(2.12), given by,  
\begin{align*}
\bar\delta\int dtdx \Delta L_{\phi}
&=\int dtdx \left\{\bar\delta  (\Delta L_{\phi})+\partial_{\lambda} (\Delta L_{\phi} \delta x^{\lambda} )\right\} \\
&=\int dt dx \left\{
+\left(-\partial_t^2(\phi \bar\delta \phi) +\partial_x^2(\phi\bar\delta\phi)\right)
+(-\partial_t(\Delta L_{\phi}\delta t) + \partial_x (\Delta L_{\phi} \delta x) 
\right\}
\end{align*}
where the identically vanishing equation of motion is taken into consideration as in eq.(2.13). 
The energy-momentum tensor is obtained as in eq.(2.14) and the components are written explicitly, 
\begin{align*}
T^t_{\ t}&= \partial_t(\phi\dot \phi) +\Delta L_{\phi} =\partial_x(\phi\phi') \  , \ \ \ 
T^t_{\ x}=-\partial_t(\phi\phi') \ , \\
T^x_{\ x}&= \partial_x(\phi \phi') -\Delta L_{\phi} =\partial_t (\phi\phi') \  , \ \ \ 
T^x_{\ t}=-\partial_x (\phi \dot\phi) .
\end{align*}
 Energy conservation is demonstrated by the direct calculation as shown in the eq.(2.15): 
\begin{align*} 
\partial_{\mu}T^{\mu}_{\ t}=\partial_t T^t_{\ t} + \partial_x T^x_{\ t} 
=\partial_t (\partial_t (\phi\dot \phi) + \Delta L_{\phi}) -\partial_x^2(\phi\dot\phi)=0
\end{align*}

This simple scalar model again shows no contribution to the equation of motion and finite contribution to the 
conserved energy-momentum. 


\end{document}